\documentclass{moriond}

\def\be{\begin{equation}}
\def\ee{\end{equation}}
\def\bea{\begin{eqnarray}}
\def\eea{\end{eqnarray}}

\newcommand{\Photo}{}

\begin{document}
\vspace*{4cm}
\title{Outer Detector of Hyper-Kamiokande}

\author{R. Shinoda, on behalf of the Hyper-Kamiokande Collaboration}

\address{Institute for Cosmic Ray Research, The University of Tokyo,\\
5-1-5 Kashiwa-no-Ha, Kashiwa, Chiba 277-8582, Japan}

\maketitle\abstracts{
Hyper-Kamiokande (HK) is the world's largest water Cherenkov ring-imaging detector, planning to start data taking in 2028.
The Outer Detector (OD) surrounds the Inner Detector and plays a critical role
in rejecting background events entering from outside, particularly cosmic-ray muons.
We report on the selection of 8~cm diameter photomultiplier tubes (PMTs) for the OD,
comparing Hamamatsu R14374 and NNVT N2031 candidates,
and present the evaluation of cosmic-ray muon background reduction performance
using a full detector simulation.
Hamamatsu PMTs were adopted for their superior in-water detection efficiency in deep-UV and stability.
The cosmic-ray muon reduction inefficiency reaches $\mathcal{O}(10^{-6})$
with OD-based cuts alone, and $\mathcal{O}(10^{-9})$ is expected when combined with
fiducial volume cuts, which is sufficiently negligible for nucleon decay
and atmospheric neutrino analyses.}

\section{Introduction}

Hyper-Kamiokande (HK) is a next-generation water Cherenkov detector,
planning to start data taking in 2028~\cite{hk_design}.
With an approximately 190~kton fiducial volume in the Inner Detector (ID),
HK will pursue a broad physics targets including nucleon decay searches,
neutrino mass hierarchy determination, and precise measurement of
the CP violating phase $\delta_{CP}$.

HK employs a two-layer detector structure.
The ID is surrounded on all sides by the Outer Detector (OD),
a 1--2~m thick water layer optically separated from the ID.
The primary role of the OD is to reject background events
entering from outside the detector.
HK is located at a depth of ${\sim}600$~m rock (1,750~m.w.e.)\ and
cosmic-ray muons constitute the dominant background at ${\sim}50$~Hz
(${\sim}4.3 \times 10^6$/day), producing PMT hits in both the ID and OD.
In contrast, signal events such as atmospheric neutrino interactions
(${\sim}80$ events per day) and nucleon decays produce hits only in the ID.
This paper reports on the OD PMT selection and the evaluation of
cosmic-ray muon reduction performance with the OD.

\section{Specifications of the Outer Detector}

The OD will be instrumented with approximately 3600 PMTs of 8~cm diameter,
each coupled with a wavelength-shifting (WLS) plate
of $30 \times 30 \times 0.7$~cm$^3$ at the edge of the photocathode.
The WLS plates contain POPOP
(1,4-bis(5-phenyl-2-oxazolyl)benzene) as a fluorescent dopant
embedded in a PMMA substrate ($n \simeq 1.49$).
They absorb deep-UV Cherenkov photons and re-emit them
at longer wavelengths (${\sim}400$~nm) where the PMT quantum efficiency (QE) is higher,
guiding the re-emitted photons to the photocathode via total internal reflection.
The walls of the OD are lined with white Tyvek sheets
serving as diffuse reflectors with an average reflectivity of 80--90\%
in the 250--700~nm wavelength range.
The average spacing between OD PMTs is ${\sim}2.5$~m,
while a typical Cherenkov ring produced in the OD has a diameter of ${\sim}2$~m.
The overall design is optimized to maximize photon detection
from background events within budget constraints.

\begin{figure}[h]
\begin{center}
\includegraphics[width=0.75\linewidth]{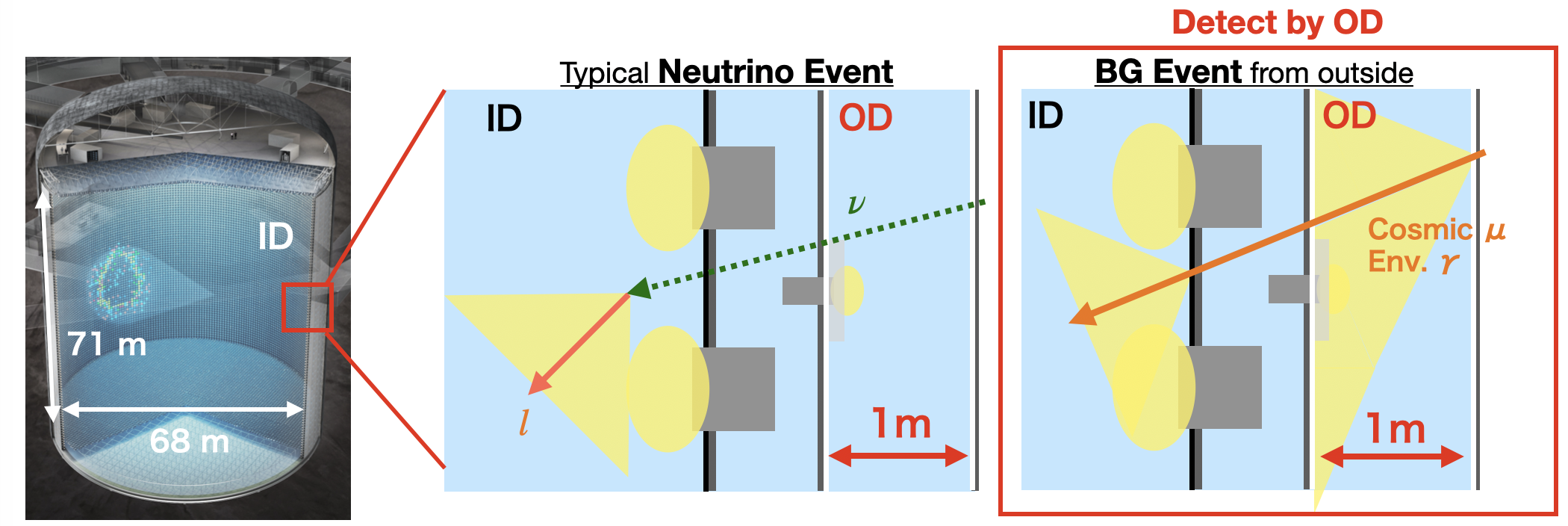}
\end{center}
\caption[]{Schematic of the typical signal and background events in the ID and OD.
Left: a typical neutrino signal event producing Cherenkov light only in the ID.
Right: a cosmic-ray muon background event detected by both the ID and OD.}
\label{fig:od_schematic}
\end{figure}

\section{OD PMT Selection}

Two candidates were evaluated for the OD PMT:
Hamamatsu R14374 and NNVT (North Night Vision Technology) N2031.
Detection efficiency and long-term stability are the critical performance
parameters for OD PMTs.

\subsection{Detection Efficiency}

Detection efficiency (DE) was measured both for the PMT alone
and in combination with an identical WLS plate submerged in water,
at four wavelengths: 405, 365, 315, and 275~nm.
A monochromatic pulsed light source at each wavelength
was directed onto either the PMT photocathode or the WLS plate individually,
and the single-photon detection rate was compared against
a calibrated reference PMT to obtain the absolute DE.
These wavelengths span from near the QE peak (${\sim}400$~nm)
down to the deep-UV region where Cherenkov radiation is intense.

For standalone PMT measurements, Hamamatsu PMTs showed
superior detection efficiency in the deep-UV region,
consistent with vendor-provided QE values within $\pm$10\% at each wavelength.
When measured through the WLS plate, however,
no significant difference was observed between the two PMT types
at any wavelength.
This is because POPOP re-emits near 400~nm where
the QE difference between the two PMTs is minimal.

Weighting the measured DE by the Cherenkov spectrum ($1/\lambda^2$),
the full-module efficiency was $35.8 \pm 1.3$\% for R14374
and $27.3 \pm 1.0$\% for N2031 giving a ratio of $1.3 \pm 0.1$ in favor of R14374.

\begin{figure}[h]
\begin{minipage}{0.48\linewidth}
\centerline{\includegraphics[width=\linewidth]{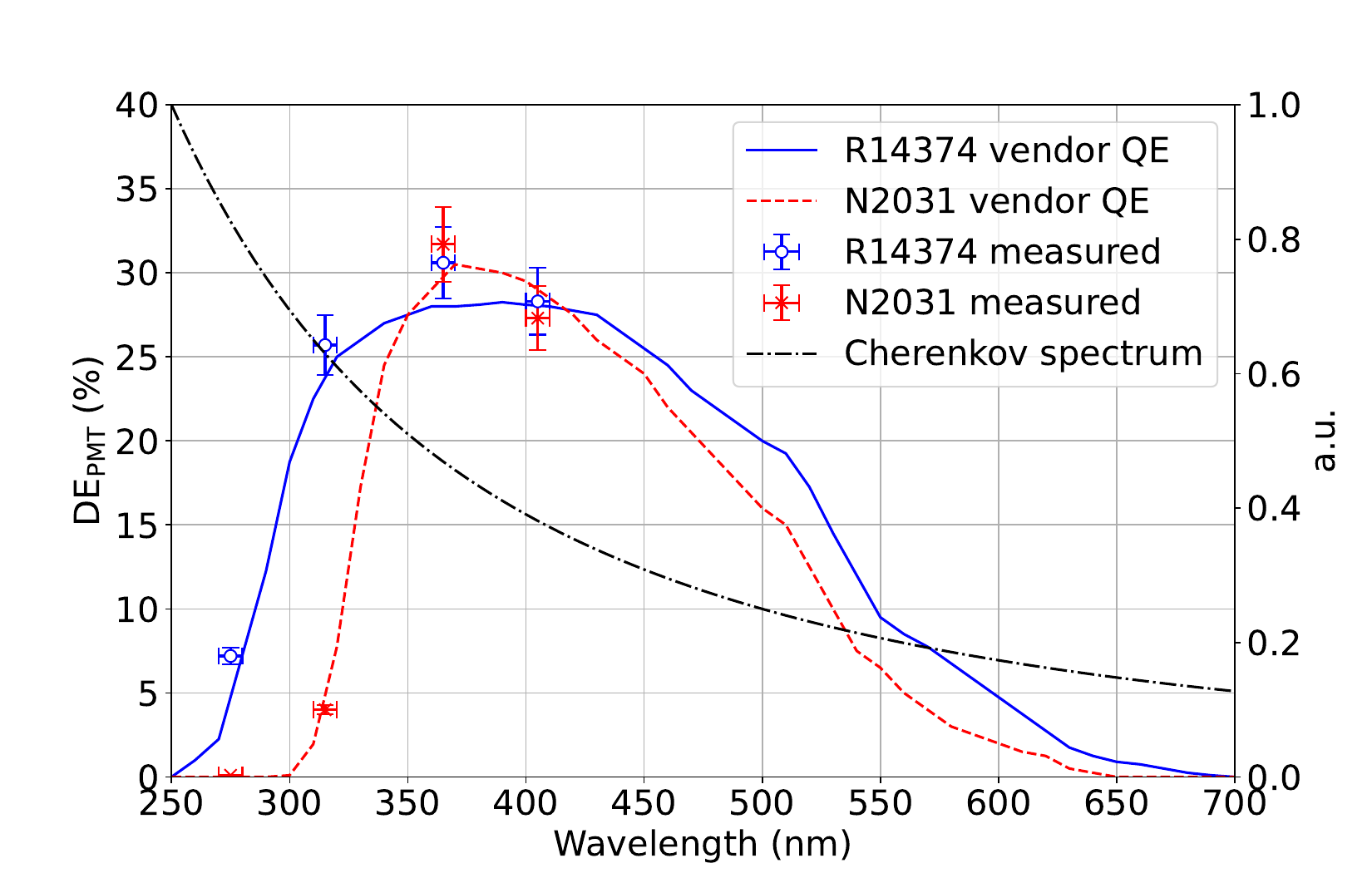}}
\end{minipage}
\hfill
\begin{minipage}{0.48\linewidth}
\centerline{\includegraphics[width=\linewidth]{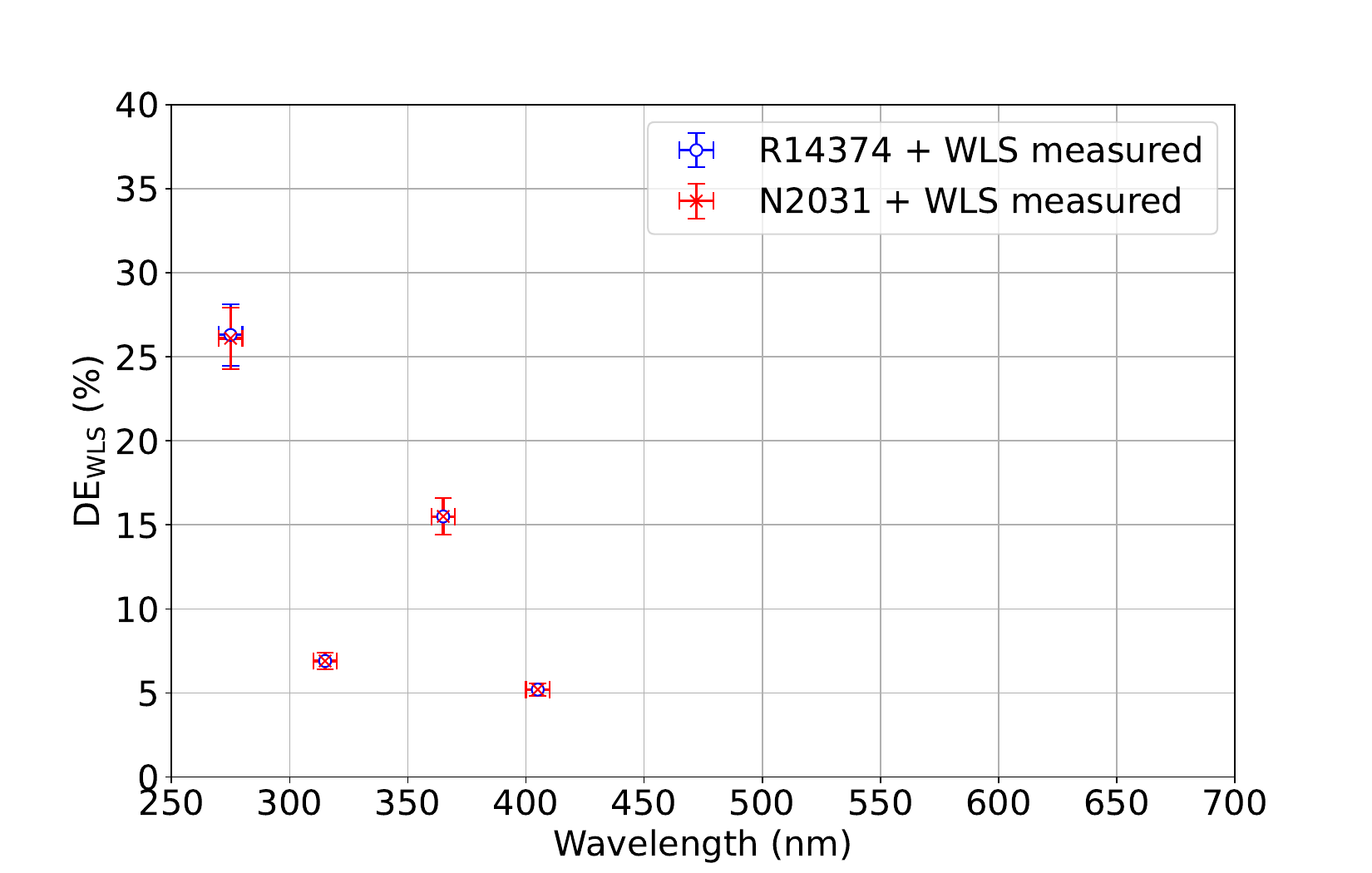}}
\end{minipage}
\caption[]{Detection efficiency at measured wavelengths.
Left: PMT standalone measurement showing Hamamatsu's superior
deep-UV efficiency.
Right: measurement through the WLS plate, where both PMTs
show comparable performance.}
\label{fig:detection_efficiency}
\end{figure}

\subsection{Basic PMT Characteristics and Stability}

Standalone PMT performances and their stability
were measured using 7 R14374 and 5 N2031 samples in air.
Evaluated parameters include: operating voltage for
a target gain of $5 \times 10^6$,
gain stability,
dark rate,
timing resolution,
and charge resolution.
The results are summarized in Table~\ref{tab:pmt_summary}.

Hamamatsu R14374 satisfied all requirements for every tested sample.
In contrast, 2 of 5 N2031 samples exceeded the 1450~V operating voltage
limit (requiring 1475~V and 1500~V),
and the N2031 mean operating voltage of 1383~V exceeded the
1300~V upper bound.
For dark rate stability, 2 of 5 N2031 samples showed
fluctuations exceeding the $\pm$0.2~kHz requirement.

Based on these comprehensive evaluations,
Hamamatsu R14374 PMTs were adopted for the HK OD.

\begin{table}[t]
\caption[]{Summary of OD PMT performance evaluation.
DE values are normalized to N2031.
Items marked with * did not meet the requirement.}
\label{tab:pmt_summary}
\vspace{0.4cm}
\begin{center}
\begin{tabular}{|l|c|c|c|}
\hline
Measured item & Requirement & R14374 & N2031 \\
 & & (satisfied/tested) & (satisfied/tested) \\
\hline
HV for $5\!\times\!10^6$ gain & 900--1450~V & 7/7 & 3/5* \\
Mean HV & 1100--1300~V & 1162~V & 1383~V* \\
Gain stability & $\pm$10\%/day & 7/7 & 5/5 \\
Dark rate & $<$2~kHz (25$^\circ$C, 0.3~p.e.)& 7/7 & 5/5 \\
Dark rate stability & $\pm$0.2~kHz (25$^\circ$C, 0.3~p.e.)& 7/7 & 3/5* \\
Time resolution & $<$3~ns FWHM & 7/7 & 5/5 \\
Charge resolution & $<$70\%~$\sigma$ & 7/7 & 5/5 \\
\hline
Cherenkov DE (est.) & --- & $1.3 \pm 0.1$ & 1 (norm.) \\
\hline
\end{tabular}
\end{center}
\end{table}

\section{Cosmic-Ray Muon Reduction Performance}

The cosmic-ray muon background reduction efficiency was evaluated
using a Geant4-based full HK detector simulation (WCSim~\cite{ref:wcsim})
incorporating measured characteristics of the Hamamatsu R14374 PMTs.
The cosmic-ray muon flux at the detector was calculated using the
MUSIC propagation code~\cite{Antonioli:1997}.
A bench-top measurement with an actual R14374 + WLS plate module
confirmed that the simulation underestimates the light yield
by ${\sim}10$\%, providing a conservative estimate.

Three sequential cuts were applied to $10^7$ equivalent
Monte Carlo cosmic-ray muon events.
The sample consists of stopping muons biased toward
low OD-hit topologies that are difficult to reject,
and was verified to encompass all such dangerous event categories
appearing in an unbiased flux sample.
The cuts exploit temporal and spatial information
of OD and ID PMT hits:

{\it 1. Typical muon reduction} (NHITAC~$<$~20):\\
Events with $\geq$20 OD PMT hits within the on-timing window
(trigger time $\pm 1~\mu$s) are rejected.
This removes the majority of cosmic-ray muon events
that produce clear Cherenkov signatures in the OD.

{\it 2. Michel electron event reduction} (NHITAC$_{1000\mathrm{ns}}$~$<$~2):\\
To identify events triggered by Michel electrons from stopped muon decays
(mean lifetime 2.2~$\mu$s),
a 1~$\mu$s sliding window is scanned from the trigger time back to $-50~\mu$s.
Using only OD PMTs within 10~m of the muon's true ID entry point,
events with $\geq$2 hits in the window with the maximum hit count are rejected.

{\it 3. Low-energy event reduction} (NHIT~$>$~166):\\
Events with on-timing ID PMT hits (NHIT) $\leq$166,
corresponding to a visible energy of ${\sim}30$~MeV,
are excluded.

\begin{figure}[h]
\begin{center}
\includegraphics[width=0.45\linewidth]{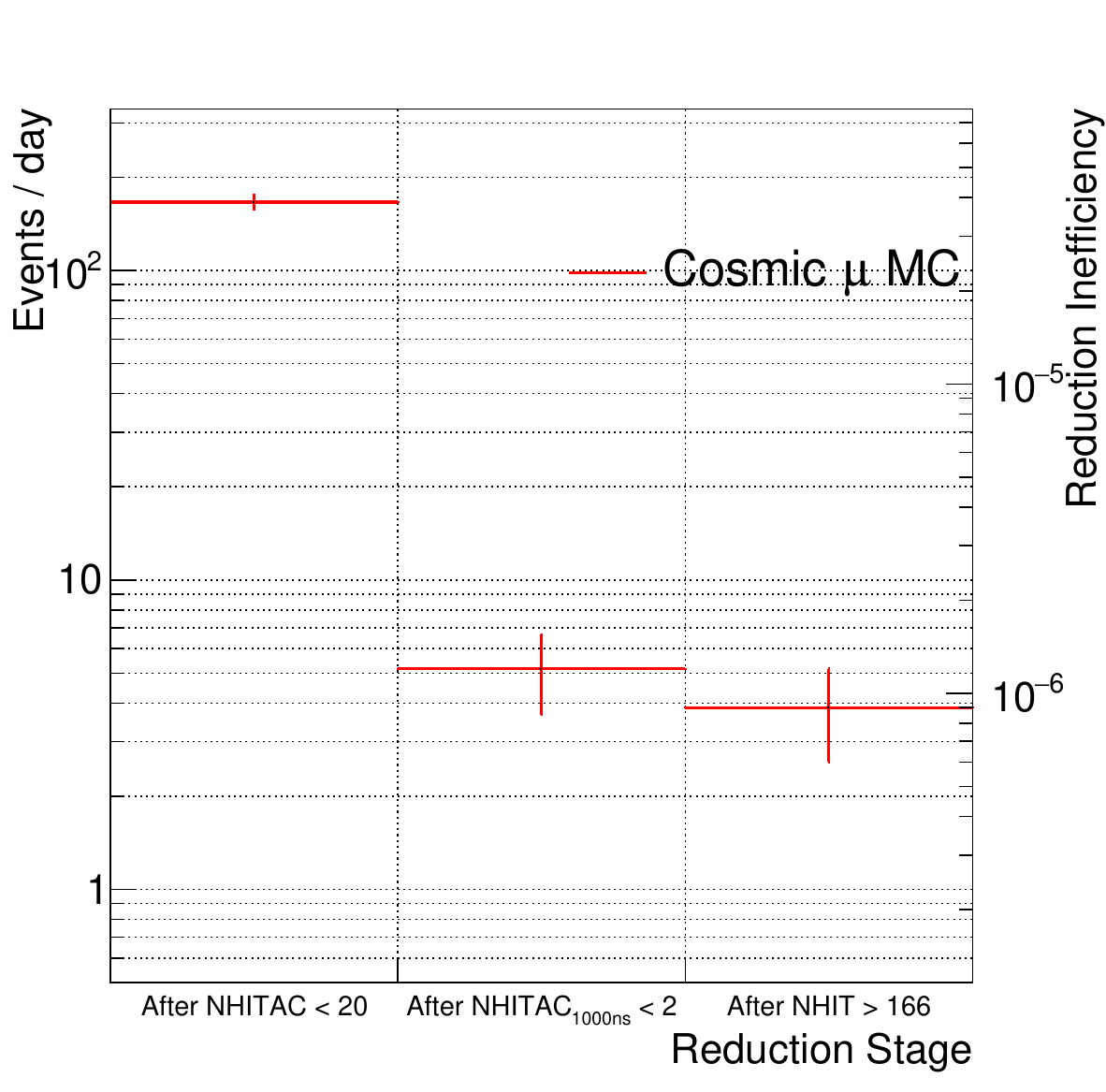}
\end{center}
\caption[]{Cosmic-ray muon events per day (left axis) and
reduction inefficiency (right axis) at each cut stage.
The reduction inefficiency reaches $\mathcal{O}(10^{-6})$
after the three-stage cuts.
Error bars represent statistical uncertainties only.}
\label{fig:reduction}
\end{figure}

After all three cuts, the reduction inefficiency reaches
$\mathcal{O}(10^{-6})$, corresponding to ${\sim}3.8$ residual
cosmic-ray muon events per day against ${\sim}80$ atmospheric
neutrino signal events per day, yielding a signal-to-background
ratio of ${\sim}20$.
The false rejection due to dark noise was also evaluated
and found to be consistent with $0 \pm 1$ event out of $10^7$.

Further improvement to $\mathcal{O}(10^{-9})$ is expected
by applying a fiducial volume cut on the reconstructed vertex.
Since cosmic-ray muons enter at the ID wall,
a fiducial volume cut of $\geq$1~m inside the wall
provides $>$3$\sigma$ separation given the ${\sim}30$~cm
vertex resolution near the wall~\cite{hk_design}.
This inefficiency is sufficiently negligible
for both nucleon decay searches and atmospheric neutrino analyses.

\section{Summary and Prospects}

Hamamatsu R14374 PMTs were selected for the HK Outer Detector,
satisfying all specification requirements with superior
detection efficiency and stability.
The cosmic-ray muon background reduction performance was evaluated
using a full detector simulation,
which conservatively underestimates the light yield by ${\sim}10$\%.
A reduction inefficiency
of $\mathcal{O}(10^{-6})$ was achieved with OD-based cuts alone, and
$\mathcal{O}(10^{-9})$ is expected when combined with fiducial volume cuts.
The OD design was confirmed to achieve the performance required
for the HK physics program.

Future work includes evaluation of long-term ($>$10~years)
detector performance, addressing degradation
from electronics and PMT malfunctions as well as aging
of the Tyvek reflectors.

\section*{References}
\bibliography{moriond}

\begin{thebibliography}{1}

\bibitem{hk_design}
K.~Abe, et al. (Hyper-Kamiokande Proto-Collaboration).
\newblock {Hyper-Kamiokande Design Report}.
\newblock 2018.
\newblock \url{https://arxiv.org/abs/1805.04163}.

\bibitem{ref:wcsim}
{The Water Cherenkov Simulator (WCSim), version 1.12.29}, 2025.
\newblock Repository:
  \url{https://github.com/WCSim/WCSim/releases/tag/v1.12.29}.

\bibitem{Antonioli:1997}
P.Antonioli, et~al.
\newblock {A Three-Dimensional Code for Muon Propagation through the Rock:
  MUSIC}.
\newblock {\em Astroparticle Physics}, 7(4):357--368, 1997.
\newblock \url{https://doi.org/10.1016/S0927-6505(97)00035-2}.

\end{thebibliography}

\end{document}